\documentclass[twocolumn,howpacs,preprintnumbers,amsmath,amssymb]{revtex4}
\usepackage{graphicx}
\usepackage{dcolumn}
\usepackage{hyperref}
\usepackage{latexsym}

\begin{document}
\title{Invasion Percolation Between two Sites}

\author{A. D. Ara\'ujo$^{1}$, T. F. Vasconcelos$^{1}$, A. A. Moreira$^{1}$, L. S. Lucena$^{2}$ 
and J. S. Andrade Jr.$^{1}$}
\affiliation{$^1$Departamento de F\'{\i}sica, Universidade Federal do
Cear\'a, 60451-970 Fortaleza, Cear\'a, Brazil.\\}
\affiliation{$^2$Departamento de F\'isica Te\'orica e Experimental, 
Universidade Federal do Rio Grande do Norte, 59072-970, Natal, RN, Brazil}

\date{\today}

\begin{abstract}
We investigate the process of invasion percolation between
two sites (injection and extraction sites) separated by a
distance $r$ in two-dimensional lattices of size $L$. Our
results for the non-trapping invasion percolation model
indicate that the statistics of the mass of invaded clusters
is significantly dependent on the local occupation
probability (pressure) $p_{e}$ at the extraction site. For
$p_{e}=0$, we show that the mass distribution of invaded
clusters $P(M)$ follows a power-law $P(M)\sim M^{-\alpha}$
for intermediate values of the mass $M$, with an exponent
$\alpha=1.39\pm0.03$. When the local pressure is set to
$p_{e}=p_{c}$, where $p_{c}$ corresponds to the site
percolation threshold of the lattice topology, the
distribution $P(M)$ still displays a scaling region, but
with an exponent $\alpha=1.02\pm0.03$. This last behavior is
consistent with previous results for the cluster statistics
in standard percolation.  In spite of these discrepancies,
the results of our simulations indicate that the fractal
dimension of the invaded cluster does not depends
significantly on the local pressure $p_{e}$ and it is
consistent with the fractal dimension values reported for
standard invasion percolation. Finally, we perform extensive
numerical simulations to determine the effect of the lattice
borders on the statistics of the invaded clusters and also
to characterize the self-organized critical behavior of the
invasion percolation process.
\end{abstract} 
\pacs{61.43.Gt,47.55.Mh,64.60.-i}
\maketitle


\section{Introduction}
Multiphase flow phenomena in porous media are relevant in
many problems of scientific and industrial importance,
including extraction of oil and gas from underground
reservoirs and transport of contaminants in soils and
aquifers \cite{Bear_72,Dullien_79}. From a theoretical point of
view, these types of processes are very complex and hence
difficult to be described in great detail. It is under this
framework that pore network models have been used to
represent the porous media and concepts of percolation
theory have been successfully applied to model slow flow of
fluids through disordered pore spaces
\cite{Feder_88,Wilkinson_83,Stauffer_94,Chandler_82,Bunde_96,
Sahimi_95,Murat_86,Tian_99,Lee_99}. For example, the method
frequently used for oil exploration consists in the
injection of water or a miscible gas (carbon dioxide or
methane) in one or more wells in the field, in order to
displace and remove oil from the interstices of the porous
rock. The extraction process persists until the interface of
separation between the two fluids (water or gas and oil)
reaches the extraction well where the injection fluid breaks
through.  At this moment, a decrease in the oil production
occurs. Because of economical interests, it is important to
determine, or at least estimate, the time when the oil
production starts to decay. It is thus natural that
companies of oil exploration demonstrate a great interest in
the knowledge and development of this displacement process,
specially in the investigation of injection dynamics, since
the time associated with it is intrinsically related to the
amount of oil that can be extracted from the underground.

The production of oil in the petroleum field must be
understood at the microscopic as well as at the macroscopic
level. During extraction, a decrease in pressure at the oil
layer near the well usually leads to the movement of water
and gas into this low-pressure zone. Ultimately, when more
distant oil cannot reach the well, water, gas, or both, are
produced instead \cite{Clark_02}. In this case, the oil
production is controlled at the macroscopic level, but
eventually microscopic processes can also limit oil
recovery. For example, capillary forces between water and
rock can effectively close off small throats in a rock
formation, so that oil cannot pass through them. The
blockage at some pore throats can therefore prevent the
extraction of oil in a large region of the oil field.

It is important to determine the relevance of the involved
forces in the displacement process. For this purpose, the
{\it capillary number} is usually defined as
\begin{equation}
Ca\equiv \frac{u\mu}{\sigma},
\end{equation}
where $u$ is the flow velocity, $\mu$ is the viscosity of
the displaced fluid and $\sigma$ is the interfacial tension
between two fluids. This dimensionless parameter gives the
relation between viscous and capillaries forces associated
to the phenomenon. For high values of $Ca$, the viscous
forces dominate the displacement dynamics. In the hypothesis
of low values of $Ca$, capillary forces dominate and the
dynamic of the displacement process is essentially
determined at the level of pores, i.e., it is intrinsically
dependent on the local aspects of the geometry of the pore
space. Under these conditions, heterogeneities of the porous
media can be characterized by a random field, representing
the spatial distribution of pore sizes, and the invasion
percolation (IP) model can be applied to study the
displacement of a fluid into a porous matrix.

The IP model and its variants have been extensively used to
simulate the process of displacement of a {\it wetting}
fluid through a porous medium by means of the injection of a
{\it nonwetting} fluid with different viscosity. Such model
has been very efficient when the injection process is
quasi-static, i.e., in the regime of low velocities. The
phenomenon of displacement is then described by the growth
of a cluster on a lattice, assuming that its border
(perimeter) represents the interface of separation between
the two fluids. Previous studies on this subject have shown
that the cluster formed during the injection process is a
fractal object with a fractal dimension close to that found
in the traditional percolation model
\cite{Wilkinson_83,Chandler_82,Roux_89}. More recent studies have
suggested that these dimensions are identical
\cite{Mark_02}. On account of its applicability, the IP model has been 
widely investigated in an attempt to determine its basic
properties and to improve the predictions for problems of
practical interest.

In this work, the non-trapping invasion percolation (NTIP)
model is applied in order to simulate the evolution of the
interface between an invading and a defending fluid in a
porous medium between two sites (wells) separated by a
distance $r$. We investigate the behavior of the mass
distribution of the invaded cluster for different values of
the pressure in the extraction site. In this way, the amount
of fluid extracted from the heterogeneous medium is
statistically quantified. The organization of the paper is
as follows. In Sec.~$2$ we briefly summarize the model used
in the simulations. In Sec.~$3$, we present the results from
computational simulations, while in Sec.~$4$ we conclude by
discussing the significance of these results and possible
applications.


\section{The Model}
In the oil extraction process, when a wetting fluid (e.g.,
water) is injected slowly into a porous medium saturated
with a nonwetting fluid (e.g., oil), capillary forces are
the major driving forces. They determine the motion of the
fluid when the injection process is carried out in the
limit where $Ca\rightarrow0$, i.e., the viscous effects are
neglected in each pore, compared with the capillary
effects. The IP model is useful in describing the basic
features of this extraction process. In this situation, the
pressure difference $\Delta p$ between the two fluids across
the meniscus follows the equation
\begin{equation}
\Delta p=\left(\frac{2\gamma}{r_{p}}\right)\cos\theta,
\label{eq_1}
\end{equation} 
where $r_p$ is the pore radius, $\gamma$ is the interfacial
tension and $\theta$ is the angle between the interface of
the two fluids and the pore wall. If we consider that
$\theta$ and $\gamma$ are constant, the advance of the
separation interface will take place through regions of lower
values of the capillary pressure, namely, those associated
with large values of the local pore radius $r_{p}$. 

In order to model the displacement of the interface between
two fluids, we use the standard NTIP model in a
two-dimensional square network. In this model, the displaced
fluid is considered to be infinitely compressible and the
injected fluid can penetrate through any region of the
interface of separation into the fluid to be
displaced. A regular lattice is adopted in the present study
as an idealization of the porous medium and we assume that
its sites represent elementary pore units. Following
Eq.~(\ref{eq_1}), the microscopic features of the disordered
porous medium are represented in this model through a random
variable $p\approx 1/r_{p}$ that gives the local
accessibility of the pore space.

Different from previous studies where the NTIP is applied
between two bars or from the center of the network to any of
its edges, here the propagation front evolves from a site
$W_{1}$ (injection site) through the network until it
reaches another site $W_{2}$ (extraction site). These two
sites $W_{1}$ and $W_{2}$ are located symmetrically between
the limits of the network and separated by a distance
$r$. Moreover, in our simulations the pressure at the
extraction site $W_{2}$ is arbitrarily specified. We will
show that the distribution of mass of the invaded clusters
is significantly dependent on this local value. For
completeness, our NTIP algorithm is described as
follows. Initially, a random value $p_{i}$ obtained from a
uniform distribution limited in the range $[0,1]$ is
distributed to each site of the network. In the initial
step, the fluid to be displaced occupies all pores of the
network, while the invading fluid is pushed through a single
site, the site $W_{1}$. We search among the neighboring
sites of $W_{1}$ the one which carries the smallest random
number $p$. This site is then invaded, becoming part of the
region occupied by the invading fluid, and the list of sites
that are eligible to be invaded next is updated. The
invasion process continues until the site $W_{2}$ is
occupied, when the mass $M$ of the invaded cluster
is measured. For a fixed value of the local pressure
at $W_{2}$, $p_{e}$, this process is repeated for several
different network realizations and the distribution of mass
of the invaded cluster $P(M)$ is calculated.

\begin{figure}
\begin{center}
\includegraphics[width=8.0cm]{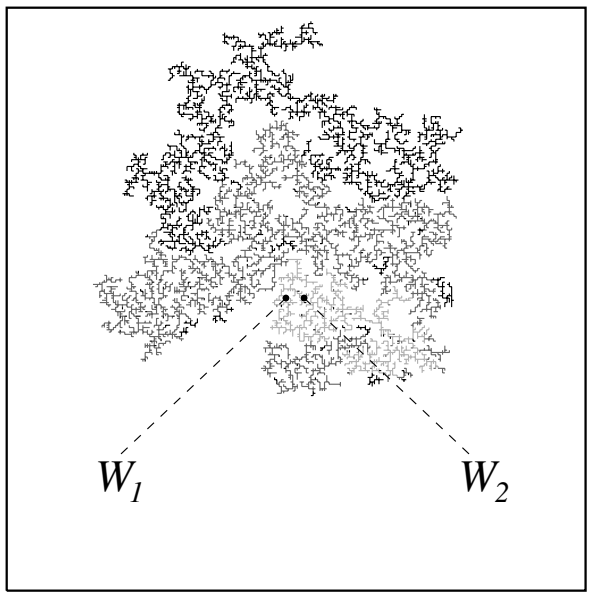}
\vspace {0.5cm}
\caption{Time evolution of the invasion process between 
the injection ($W_{1}$) and extraction ($W_{2}$) sites. The
steps of invasion are identified in accordance with the
number of iterations (time) executed during the simulation
process. The colors indicate in which stage of the process
the propagation front reached a given site. The light gray
sites correspond to the range $1\leqslant t\leqslant 1000$,
the dark gray sites to $1000 < t\leqslant 5000$, and the
black ones correspond to $t > 5000$. The pressure value at
the extraction well is $p_{e}=0$.}
\label{fig_1}
\end{center}
\end{figure}

\begin{figure}
\begin{center}
\includegraphics[width=8.0cm]{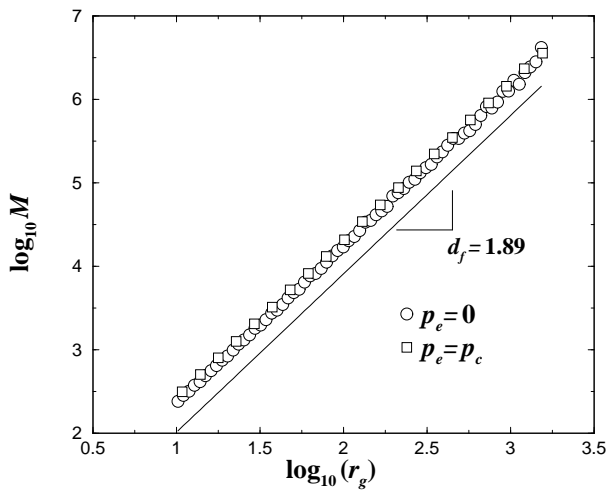}
\caption{Log-log plot of the radius of gyration for $L=8192$ and $r=4$. 
The circles correspond to $p_{e}=0$ and the squares to
$p_{e}=p_{c}$. These results have been produced with
$10000$ realizations of NTIP clusters that do not touch the border
of the lattice. The straight line (shifted downwards for
better visualization) is the least-square fit
to the data and the slope gives the fractal dimension
$d_{f}=1.89\pm0.01$ .}
\label{fig_2}
\end{center}
\end{figure}
%

\section{Results and Discussion}

In Fig.~\ref{fig_1}, we illustrate some stages of the
invasion process taking place in a lattice of size $L=256$
and distance between sites $r=16$ . The shades of gray
indicate in which interval of time of the process a given
site has been invaded. As can be seen from this typical
realization of the model, the cluster resulting from the
invasion process represents only a small fraction of the
network.

In what follows, we show how sensitive is the statistics of
the mass of the invaded clusters to the pressure value
$p_{e}$ imposed at the extraction site $W_{2}$. Since the
invasion finishes as soon as the extraction well is reached,
we can interpret that the local pressure $p_e$ corresponds to a
special criteria to stop the NTIP growth process, without
any interference in the microscopic rules of the model.
This suggests that the invaded clusters obtained for
distinct values of $p_e$ should still be in the same
universality class observed for invasion percolation between
two borders of a lattice. A direct evidence for this
fact can be obtained through the calculation of the fractal
dimension of the invaded cluster. We show in
Fig.~\ref{fig_2} the log-log plot of the mass of the invaded
cluster against its radius of gyration for $p_{e}=0$ and
$p_{e}=p_{c}$. These results have been obtained for,
$L=8192$, $r=4$ and $10000$ realizations in which the growth
process stopped before the invading cluster reached the
border of the lattice. The best linear fit to both data sets
gives approximately the same value for the fractal dimension
$d_{f}=1.89 \pm 0.01$. This value is consistent with the
value reported for the standard invasion percolation
process, $d_{f}\approx 1.89$
\cite{Feder_88,Stauffer_94,Schwarzer_99,Knackstedt_02}. Our
results therefore support the assumption that NTIP and SP
are in the same universality class \cite{Stauffer_94}.

\begin{figure}
\begin{center}
\includegraphics[width=8.0cm]{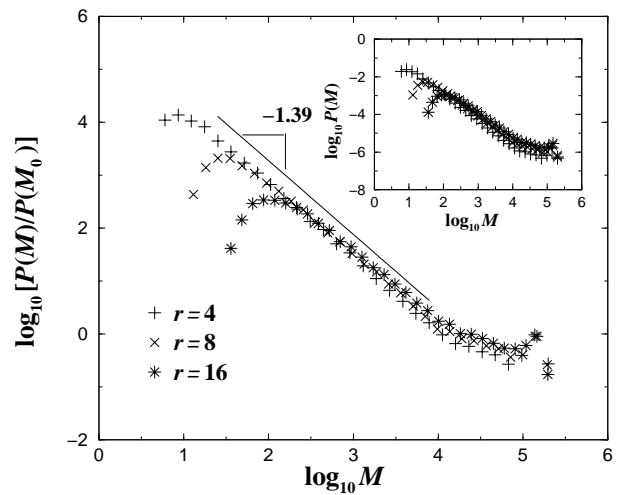}
\caption{Graph in log-log scale of the mass distribution 
of invaded clusters $P(M)$ for the case in which the
pressure at the extraction site is $p_{e}=0$, and a network
size $L=512$. Each curve correspond to a different value for
the distance between the wells, namely $r=4$ (plus), $8$ (X)
and $16$ (stars). These distributions have been rescaled by
its corresponding value $M_{0}$ at the position of the
bump. The straight line is the least-square fit to the data
in the scaling region with the slope ~$-1.39\pm
0.03$. For better visualization the straight line has been
slightly shifted upwards, i.e., the ordinate values have
been multiplied by an arbitrary constant factor. The
corresponding unscaled distributions are shown in inset.}
\label{fig_3}
\end{center}
\end{figure}

In Fig.~\ref{fig_3} we show the log-log plot of the
distribution $P(M)$ of invaded masses for 
the case $p_e=0$, and
different values of the ``well'' distance $r$. Clearly, all
distributions displays power-law behavior $P(M)\sim
M^{-\alpha}$ for intermediate mass values. In addition, a
lower cutoff of order $r^{d_f}$ is always present,
corresponding to the minimum cluster size. The upper cutoff
of order $L^2$, on its turn, refers to the situation in
which almost all sites of the lattice are invaded before the
invasion front reaches the extraction site. The effect of
$r$ on the distribution is to simply modify the range of the
scaling region. The solid line in Fig.~\ref{fig_3}
corresponds to the best linear fit for the data
corresponding to $r=4$ in the scaling region, with the
exponent $\alpha=1.39\pm0.03$. These curves also present a
``bump'' in the region of large values of mass. For
comparison, when we rescale the distributions by their
corresponding value at the maximum of this bump $M_{0}$, a
collapse in the scaling region can also be observed.
\begin{figure}
\begin{center}
\includegraphics[width=8.0cm]{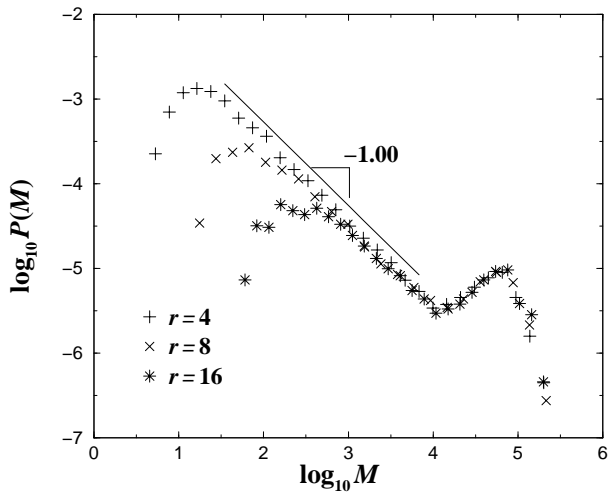}
\caption{Log-log plot of  $P(M)$ for the case $L=256$ and
different values for the distance between the wells, $r=4$
(plus), $r=8$ (X), and $r=16$ (star). The
pressure at the extraction site is equal to
$p_{e}=p_{c}\simeq 0.5927$. The solid line indicates the
linear regression (for $r=4$) with the exponent $\alpha=1.00\pm
0.03$. For better visualization, it has been slightly
shifted upwards. }
\label{fig_4}
\end{center}
\end{figure}
\begin{figure}
\begin{center}
\includegraphics[width=8.0cm]{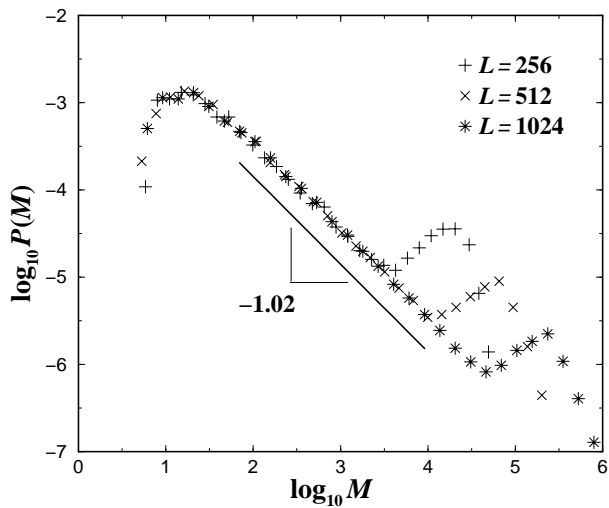}
\caption{Log-log plot of  $P(M)$ for the case $r=4$ and
different lattice sizes, $L=256$ (plus), 512 (X), and 1024
(star). The pressure at the extraction site is equal to
$p_{e}=p_{c}\simeq 0.5927$. The solid line indicates the
linear regression (for $L=1024$) with the exponent
$\alpha=1.02\pm 0.03$.}
\label{fig_5}
\end{center}
\end{figure}

In Fig.~\ref{fig_4} we show the distributions of mass $P(M)$
for the case in which the pressure at the site $W_{2}$ is
arbitrarily set to $p_{e}=p_{c}\simeq0.5927$, where $p_{c}$
corresponds to the site percolation threshold of the square
network \cite{Stauffer_94}. Again, the smaller the distance
between the sites, larger is the range over which the
power-law behavior holds. In addition, the distributions
present the same lower cutoff of order $r^{d_f}$ and the
upper cutoff of order $L^{2}$, where a bump can also be
observed in the region of large values of mass
\cite{Barth_99,Araujo_03}.  We show in Fig.~\ref{fig_5} 
the effect of the system size on the distributions of mass
for the case $r=4$ and $p_e=p_c$. The results shown in
Figs.~\ref{fig_4} and \ref{fig_5} are also consistent with
the existence of a power-law regime for intermediate values
of mass, $P(M)\sim M^{-\alpha}$, but with an exponent
$\alpha=1.02\pm 0.03$. The less negative exponent in the
power-law distribution for the cases $p_e=p_c$ when compared
to $p_e<p_c$ means that, in the later case, larger clusters
appear more frequently.

As already mentioned, just before the extraction site is
invaded, all sites belonging to the invasion front have
their pressure values larger than $p_e$. Therefore, in the
case where $p_{e}=p_{c}$, the invaded cluster is similar to
a cluster generated using the standard percolation (SP)
model at the critical point. In the SP model the
distribution of cluster sizes follows a power-law, $P(S)\sim
S^{-\tau}$, where the exponent $\tau\approx 2.055$ is known
as {\it Fisher exponent} \cite{Stauffer_94}. Assuming that
the IP and SP models are compatible, note that, in our model
the invaded cluster would be the same if any of its sites
were chosen as the injection point. As a consequence, the
probability of finding an invaded cluster of size $M$ is the
product of two factors, namely, the probability with which
such cluster appears in a percolation lattice at the
critical point, and the size of the cluster. Thus, we should
expect $\alpha=\tau -1\approx 1.055$, a value that is in
good agreement with our numerical results.

\begin{figure}
\begin{center}
\includegraphics[width=8.0cm]{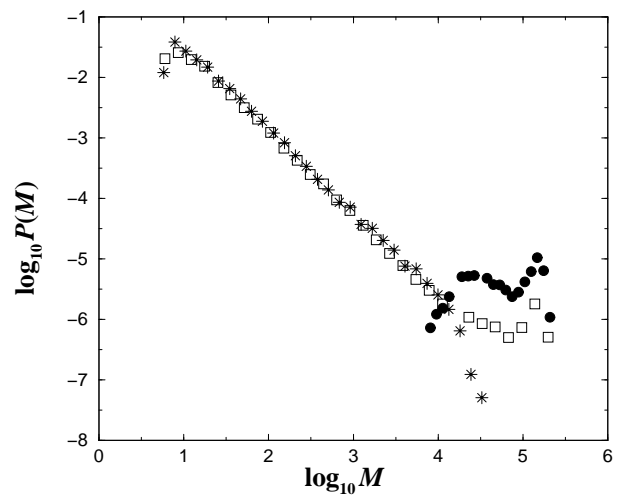}
\caption{Log-log plot of the distribution $P(M)$ for
$p_{e}=0$ and $L=512$. The squares include all clusters, the
stars only the clusters that do not touch any of the edges,
and the full circles only those that touch the edges of the
lattice substrate. All distributions have been normalized by
their integrals.}
\label{fig_6}
\end{center}
\end{figure}
\begin{figure}
\begin{center}
\includegraphics[width=8.0cm]{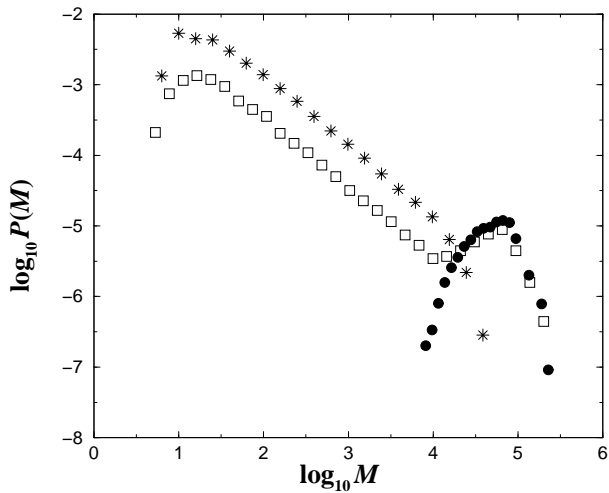}
\caption{Log-log plot of the distribution $P(M)$ of the 
invaded cluster $p_{e}=p_{c}$ and $L=512$. The squares
include all clusters, the stars only the clusters that do
not touch any of the edges, and the full circles only those
that touch the edges of the network. All distributions have
been normalized by their integrals.}
\label{fig_7}
\end{center}
\end{figure}

\begin{figure}
\begin{center}
\includegraphics[width=8.0cm]{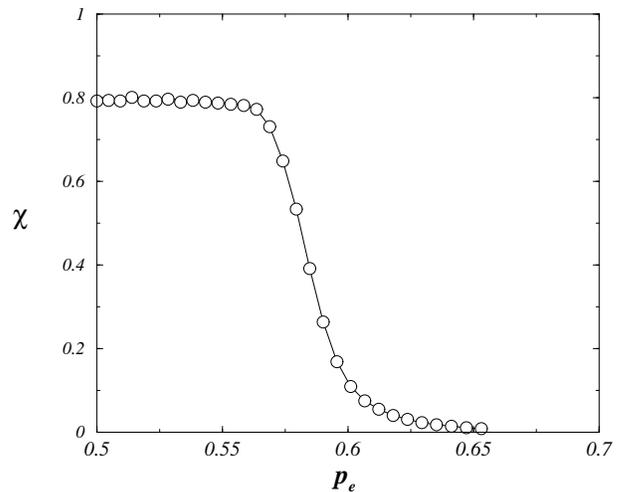}
\caption{The fraction $\chi$ of invading clusters 
that do not touch the edges of the network as a function of the
pressure value at the extraction site for system size
$L=256$ and distance between sites $r=8$.}
\label{fig_8}
\end{center}
\end{figure}

The bumps present in the distributions shown in
Figs.~\ref{fig_3} and \ref{fig_4} are due to aggregates that
grow long enough to reach the border of the lattice before
reaching the extraction well.  This is made clear when we
distinguish these clusters in our simulation data from those
that stop before reaching the borders. The distributions
shown in Fig.~\ref{fig_6} and Fig.~\ref{fig_7} reveal that
the bumps are mainly originated by the cluster masses
subjected to size limitations (full circles), i.e., those
clusters that touch the lattice borders. Note that, even
after touching the border, the cluster continues to grow
until the extraction well is invaded. Thus, those clusters
that contribute to the bump in the region of large $M$ have
a fractal dimension $d_f=2$ and therefore do not possess the
critical properties observed in percolation clusters. The
statistics of these larger clusters can be investigated
if we compute the fraction $\chi$ of invaded clusters that
do not touch the edges of the simulation box for different
values of the pressure $p_{e}$. As shown in
Fig.~\ref{fig_8}, for a given value of $L$ and $r$, $\chi$
remains approximately constant up to a value of $p\approx
p_{c}$ where it experiences a sharp decrease and reaches
$\chi=0$. For $p_e<p_c$, the presence of these large
clusters is an artifact of the finite size of the lattice
and becomes less relevant as the lattice size grows. In the
results which follow we will only consider the clusters that do
not touch the borders when computing the distributions of
mass.

\begin{figure}
\begin{center}
\includegraphics[width=8.0cm]{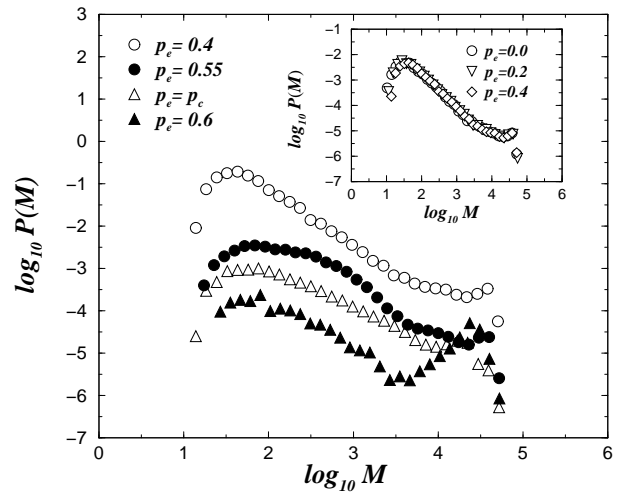}
\caption{Log-log plot of the mass distribution of 
the invaded cluster for different values of the pressure at
the extraction site. $p_{e}=0.4$ (circles), $p_{e}=0.55$
(full circles), $p_{e}=p_{c}$ (triangles) and $p_{e}=0.6$
(full triangles). The inset shows the mass distribution of
invaded cluster for pressure values $p_{e}=0.0$ (circles),
$p_{e}=0.2$ (triangles) and $p_{e}=0.4$ (diamond). The
system size is $L=256$ and the distance between the
injection and extraction sites is $r=8$.}
\label{fig_9}
\end{center}
\end{figure}

To better understand how the statistics of the mass
distribution changes with the value of the pressure at the
extraction site, we perform simulations with different
values of $p_{e}=0.0,~0.2,~0.4,~0.55, ~p_{c}~(\simeq
0.5927)$, and $0.6$. From the results presented in
Fig.~\ref{fig_9}, we observe that the distribution $P(M)$
remains practically unchanged (see the inset) with a scaling
regime at intermediate values of $M$ that persists at least
up to $p_{e}=0.4$. For $p_{e}$ values larger than this, the
distribution $P(M)$ changes gradually (see the curve for
$p_{e}=0.55$) until another power-law is effectively
established at $p_{e}=p_{c}$. For $p_{e}>p_{c}$, most of the
generated clusters are subjected to border effects and the
scaling behavior is again suppressed. 

\begin{figure}[t]
\begin{center}
\includegraphics[width=8.0cm]{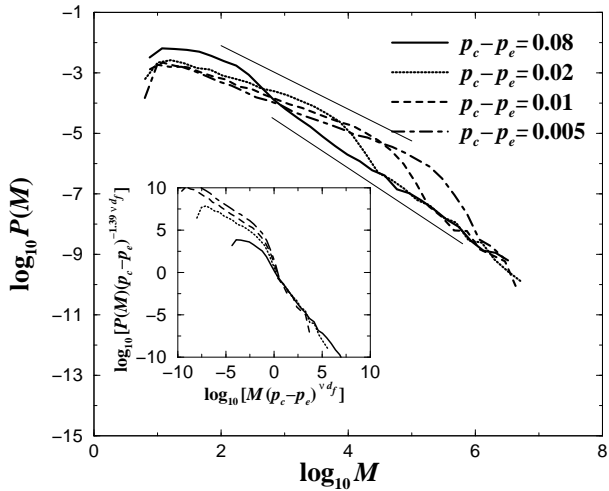}
\caption{
Log-log plot of the mass distribution of the invaded cluster
for values of the pressure at the extraction site
approaching the critical point $p_e=p_c$ for systems size
$L=8192$. The thin lines are included as guide to eyes and
have the slopes expected in the following limiting cases:
({\it{i}}) $-1.05$ for the critical case $p_e=p_c$ and
({\it{ii}}) $-1.39$ observed in $p_e=0$. In the inset,
the curves are rescaled by the characteristic mass
$\xi^{d_f}$, where $\xi$ is the correlation coefficient which
scales with the pressure at the extraction well as $\xi \sim
(p_c-p_e)^{-\nu}$.  The exponents $d_f$ and $\nu$ are
assumed to be the same as in the the standard percolation
model, namely, $d_f=1.89$ and $\nu=4/3$.}
\label{fig_10}
\end{center}
\end{figure}

The behavior of the distributions as $p_e$ approaches $p_c$
is shown in Fig.~\ref{fig_10}. For large values of $M$, all
curves converge to the same behavior, which coincides with
the power-law form obtained for the case $p_e=0$. For small
values of $M$, the curves display a power-law decay with an
exponent that depends on the value of $p_e$, but approaches
the value $\alpha=1.055$, expected in the limit
$p_e=p_c$. One can understand this behavior by making a
parallel with the way in which the standard percolation
model approaches the critical point. As $p \to p_c$ in SP,
the characteristic length of the clusters $\xi$ diverges as
$(p_c-p)^{-\nu}$, with $\nu=4/3$. Thus, we should observe
the critical behavior for values of $M$ bellow the
characteristic mass of a cluster obtained with $p=p_e$, and
the off-critical behavior for values of $M$ above this
characteristic mass. It means that the onset of the
crossover between the two behaviors should scale as
$(p_c-p_e)^{-\gamma}$, with $\gamma=\nu d_f$. Assuming that
our invasion process is consistent with SP, we expect
$\gamma\approx 2.52$. This {\it ansatz} is supported by the
scaled curves shown in the inset of Fig.~\ref{fig_10}. As
can be seen, for $p_e \ll p_c$, the crossover moves to
values of mass smaller than the cutoff region in $r^{d_f}$,
and only the off-critical regime is present. On the
other hand, for $p_{e}>p_{c}$, most of the generated
clusters reach the border of the lattice, leaving the
critical regime, and the scaling behavior is again
suppressed.

It is possible to understand why we obtain distinct
distributions in the critical ($p_e=p_c$) and off-critical
($p_e<p_c$) regimes by noting that our invasion percolation
process can be divided in the following two
sequential phases. In the first phase, the invaded cluster
grows from the injection site and reaches an immediate
neighborhood of the extraction site. The second phase
includes all additional invading steps up to the point in
which the extraction site itself is taken, and the invasion
process is then terminated. Thus, we can say that the total
invaded mass is the sum of two variables, $M=M_1+M_2$, where
$M_1$ and $M_2$ are the numbers of invasion steps to
accomplish the first and second phases, respectively. Since
the masses $M_1$ and $M_2$ have distinct distributions,
namely, $P_1(M_1)$ and $P_2(M_2)$, respectively, the
distribution with the longer tail between them, should
determine the behavior of the tail of $P(M)$. Note that the
distribution $P_1(M_1)$ does not depend on the value of
$p_e$. On other hand, the number of additional steps to
invade the extraction well clearly depends on $p_e$ and so
does $P_2(M_2)$. In the particular case of $p_e=0$, it follows
that $M_2=1$. We can then write that $P(M)=P_1(M-1) \simeq
M^{-\alpha}$, which shows that the exponent $\alpha=1.39$
comes from the distribution of steps necessary to complete
phase 1. When $p_e$ approaches $p_c$, the contribution of
$M_2$ to the distribution $P(M)$ starts to become
relevant. In the critical condition $p_e=p_c$, the tail of
this distribution is dominated by the number $M_2$ of steps
to invade the extraction well. Although we lack information
about the complete form of $P_2(M_2)$, we know that the
necessary condition to invade the extraction well is that
all sites belonging to the perimeter of the invaded cluster
have pressures $p>p_e$. Following percolation theory
\cite{Stauffer_94}, when $p_e=p_c$, this condition leads to a
power-law distribution, $P(M) \simeq M^{-\alpha}$, where
$\alpha=1.055$. As a consequence, we can deduct that, to be 
dominant, the tail of $P_2(M_2)$ should behave in the same 
way.


\begin{figure}[t]
\begin{center}
\includegraphics[width=8.0cm]{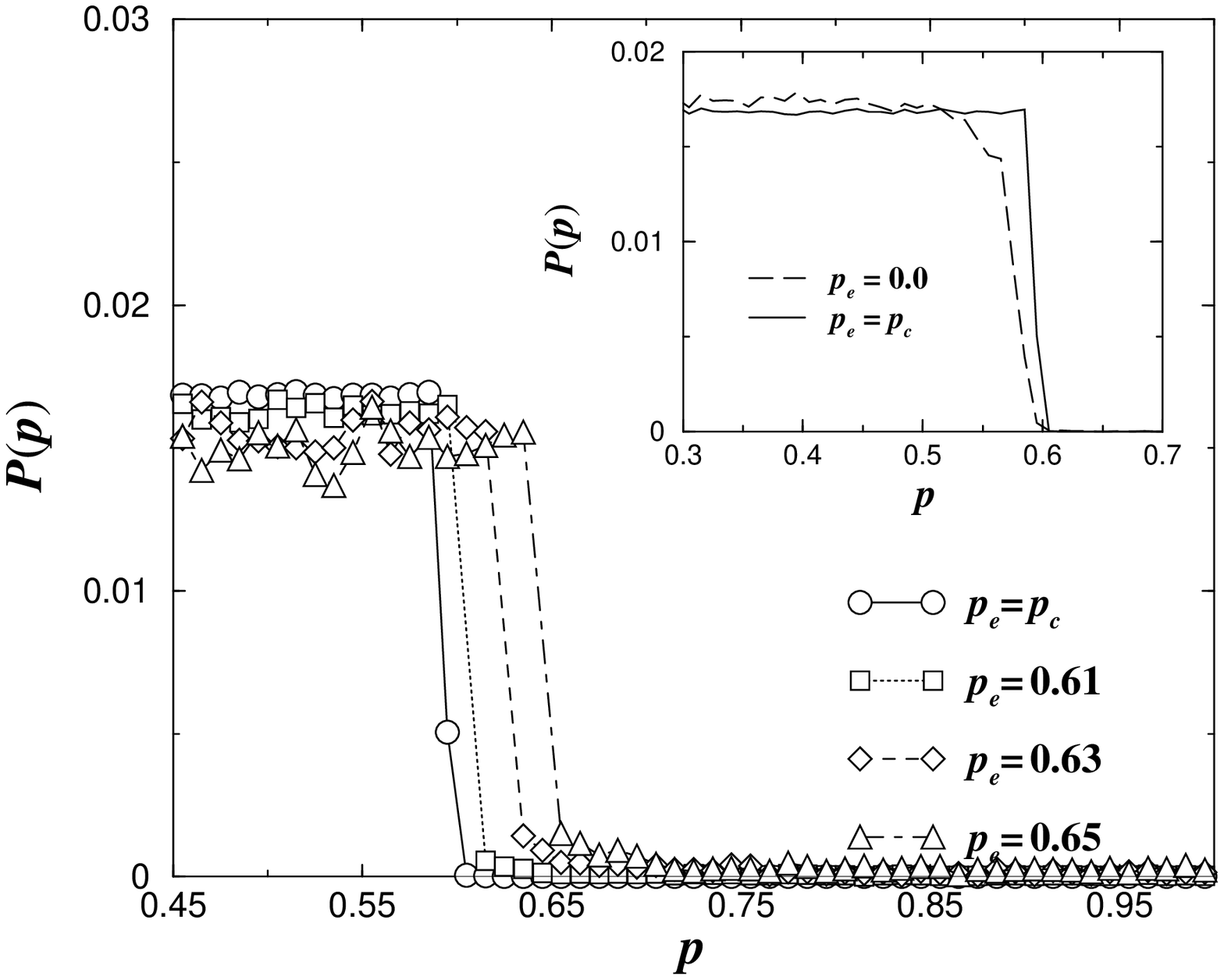}
\caption{The probability distribution $P(p)$ of invaded sites for
different values of the pressure in the extraction site,
$p_{e}=p_{c}$ (circles), $0.61$ (squares) and $0.63$
(diamonds) and $0.65$ (triangles) for $L=512$ and $r=4$. The
inset shows the probability distribution $P(p)$ for
$p_{e}=0.0$ (dotted line) and $p_{c}$ (solid line).}
\label{fig_11}
\end{center}
\end{figure}

The invasion percolation is broadly recognized as the most
convincing model for a natural phenomenon that displays
self-organized critical (SOC) behavior
\cite{P_Back_88}. Indeed, this characteristic is manifested,
for example, in the distribution of the random numbers
$P(p)$ corresponding to those sites that have been occupied
during the invasion process. Accordingly, the fact that
$P(p)$ displays a sharp cut-off on the right side at the
critical percolation point $p_{c}$ represents a sound
signature of the SOC behavior, which dictates the dynamics of
the displacement process. In Fig.~\ref{fig_11} we plot the
distribution $P(p)$ computed from $1000$ realizations of
size $L=512$ and distance $r=4$, and for values of $p_{e}=0,
0.61, 0.63$ and $0.65$ as well as for $p_{e}=p_{c}$. For
$p_{e}=0$ and $p_{e}=p_{c}$, after the completion of the
invasion process, both distributions display a clear cutoff
at $p\approx p_{c}$, in agreement with previous simulations
of the standard invasion percolation model
\cite{Zara_99,Araujo_04}. The inset in Fig.~\ref{fig_11} shows that the 
transition is clearly sharper for $p_{e}=p_{c}$ than for
$p_{e}=0$. In the cases in which $p_{e}>p_{c}$, we observe
that the cutoff of the distribution $P(p)$ shifts rightwards
to $p=p_{e}$. Since we are only interested in those clusters
that do not touch the borders, for $p_{e}>p_{c}$, a very
large number of trials of the invasion process is necessary
to generate the number of realizations required to produce
good statistics (see Fig.~\ref{fig_8}). Once more, if we
argue that $p_{e}$ can be interpreted as a stopping criteria
for the invasion process, the non-invaded sites that are
neighbors of the perimeter sites of clusters generated under
such constraint should all have local probabilities that obey
the inequality $p>p_{e}$. As a consequence, for
$p_{e}>p_{c}$, the invaded clusters are limited by a ``high
energy'' border of non-invaded sites.

%


\section{Conclusions}

In summary, we investigate here the statistics of the mass
of the invasion percolation clusters generated between two
sites separated by a distance $r$ in a two-dimensional
network, and for different values of the local probability
at the extraction site, $p_{e}$. The results of our
simulations indicate that the distribution of mass of the
invaded cluster $P(M)$ is highly sensitive to $p_{e}$. More
important, $P(M)$ displays power-law behavior in the
following two different conditions: ({\it i}) an
off-critical regime for sufficiently low values of $p_e$,
and ({\it ii}) a critical regime for
$p_{e}=p_{c}$. Interestingly, even though the scaling
exponents calculated in both cases are substantially
different, namely $\alpha=1.39 \pm 0.03$ and $1.02 \pm 0.03$
for cases ({\it{i}}) and ({\it{ii}}), respectively, our
results for the fractal dimension of the invaded clusters
show that they should belong to the same universality class.
For $p_e$ approaching $p_c$, we found that the distribution
displays a crossover from the critical to the off-critical
regime and that the onset of the crossover scales as
$\xi^{d_f}$, where $\xi$ is the correlation length, $\xi
\sim (p_c-p_e)^{-\nu}$.  Finally, we have shown that the SOC
behavior of the invasion percolation process is rather
robust, even in the situation in which $p_{e}>p_{c}$. In
this case, although rarely found, those clusters whose
growth has not been affected by the borders of the NTIP
system (named here ``critical clusters'') present a very
peculiar property, namely, that {\it all} the non-invaded
sites at their perimeters have an anomalously high local
probability.

\section{Acknowledgments}
We acknowledge  CNPq (CT-PETRO/CNPq), CAPES,
FINEP and FUNCAP for financial support.

\end{document}